\documentclass[%
 reprint,
 amsmath,amssymb,
 aps,
]{revtex4-1}

\usepackage{graphicx}
\usepackage{dcolumn}
\usepackage{bm}

\begin{document}

\preprint{APS/123-QED}

\title{Generating the Log Law of the Wall with Superposition of Standing Waves}

\author{Chien-chia Liu}
 \email{chien-chia.liu@ds.mpg.de}
 \affiliation{%
 Max Planck Institute for Dynamics and Self-Organization,\\
 Am Fa{\ss}berg 17, D-37077 G{\"o}ttingen, Germany
}%

\date{\today}

\begin{abstract}
Turbulence remains an unsolved multidisciplinary science problem. As one of the most well-known examples in turbulent flows, knowledge of the logarithmic mean velocity profile (MVP), so called the log law of the wall, plays an important role everywhere turbulent flow meets the solid wall, such as fluids in any kind of channels, skin friction of all types of transportations, the atmospheric wind on a planetary ground, and the oceanic current on the seabed. However, the mechanism of how this log-law MVP is formed under the multiscale nature of turbulent shears remains one of the greatest interests of turbulence puzzles. To untangle the multiscale coupling of turbulent shear stresses, we explore for a known fundamental tool in physics. Here we present how to reproduce the log-law MVP with the even harmonic modes of fixed-end standing waves. We find that when these harmonic waves of same magnitude are considered as the multiscale turbulent shear stresses, the wave envelope of their superposition simulates the mean shear stress profile of the wall-bounded flow. It implies that the log-law MVP is not expectedly related to the turbulent scales in the inertial subrange associated with the Kolmogorov energy cascade, revealing the dissipative nature of all scales involved. The MVP with reduced harmonic modes also shows promising connection to the understanding of flow transition to turbulence. The finding here suggests the simple harmonic waves as good agents to help unravel the complex turbulent dynamics in wall-bounded flow. 
\end{abstract}

\maketitle


Flows in nature are turbulent flows in general. Wall-bounded flow where flow over a solid surface is considered as a classic example. Owing to its practical importance in transportation and meteorology, wall-bounded flow is densely studied, focusing on a thin layer almost right next to the wall. Within such thin layer, the interaction between turbulence and the wall is most crucial, and the mean velocity profile (MVP) reveals the log-law behavior, so called the log law of the wall \citep{TnL,Hinze,Schlichting,Tritton,LL,Pope,McComb,GFD,AFD,GG2010}. However, the formation mechanism of the log-law MVP is still one of the greatest challenges in turbulent flows \citep{WBF2010,WBFScience,WBFNature}. It has long been postulated that the turbulent scales, which play crucial roles in forming the log-law MVP, reside in the inertial subrange associated with the Kolmogorov energy cascade, where viscosity is ineffective. Evidence of this argument is then the simultaneous observation of the log-law MVP and the Kolmogorov spectrum, which is unfortunately not at all apparent. It might help further our understanding of the log-law MVP with a new yet simple physical perspective on both how the log-law MVP may be formed and what the constants involved in the log law may signify. Thanks to the continuous success in identifying the unstable traveling waves in the non-stationary wall-bounded mean flows \citep{TWPRL,TWScience,TW}, it gives rise to the thinking of whether there exists some simple harmonic waves in the stationary wall-bounded mean flow. Here we report how to generate such log-law MVP whose mechanism is unknown using the even harmonic modes of fixed-end standing waves with the known mechanism.

\begin{figure*}
\includegraphics[width=0.9\textwidth, clip, trim=0 100 0 100, angle=0]{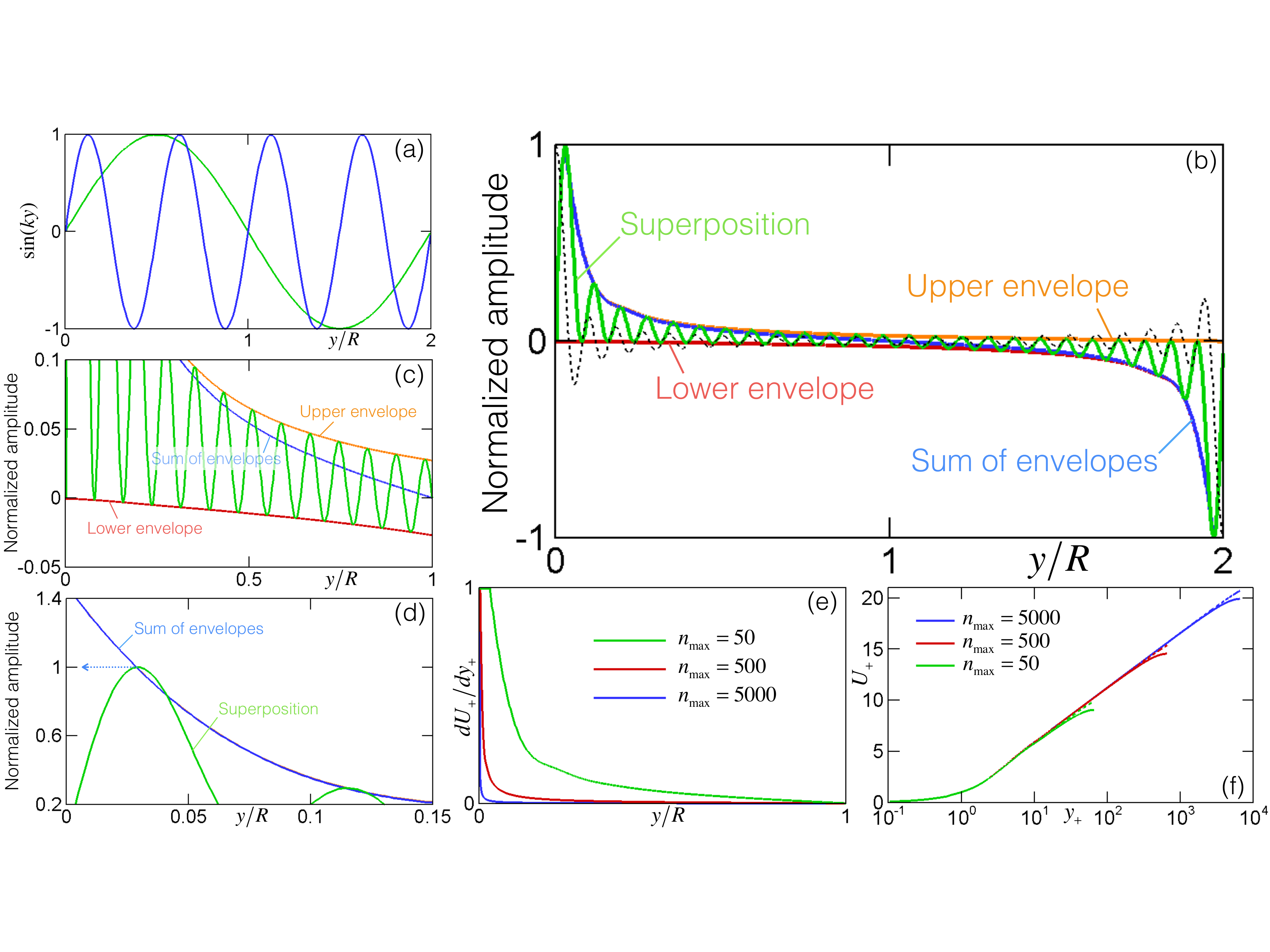}
\caption{\label{fig:epsart} 
(a) Two select examples of fixed-end standing waves with the even harmonic mode of $n_{min} = 2$ and $n = 8$. (b) The superposition of the standing waves together with its upper and lower envelopes as well as the sum of the two envelopes, using Eq. (7) with $n_{max} = 50$. The amplitude of all curves is normalized by the maximum peak value of wave superposition. For comparison, we also show the superposition (thin dotted line) of free-end standing waves with odd harmonic modes. (c) The upper envelope does not decay to zero while the lower envelope deviates equally yet negatively from zero, when both envelopes approach $y/R = 1$. To be consistent with the boundary condition where $dU_+/dy_+ = 0$ at $y/R = 1$, herein the wave envelope for further calculation takes the sum of upper and lower envelopes. (d) The wave envelope overshoots the maximum peak of wave superposition when approaching $y/R = 0$, violating the boundary condition of $dU_+/dy_+ = 1$ at $y/R = 0$. Here we simply take the amplitude of the wave envelope to be unity for $y/R$ between 0 and such maximum peak. (e) Wave envelopes or the equivalent $dU_+/dy_+$ for $n_{max}=$ 50, 500, and 5000. (f) Solid lines: MVP's determined using the results in (e) with Eq. (6-8) and $\kappa = 0.42$. Dashed lines: MVP's with the same $n_{max}$ but including both odd and even modes (from $n = 1$ to $n_{max}$) of fixed-end standing waves.  
}
\end{figure*} 

We start from the stress balance for stationary incompressible channel flow \citep{Boussinesq}
\begin{eqnarray}
(\nu + \nu_T)dU/dy = u_F ^2 (1-y/R),
\end{eqnarray}
where $\nu$ and $\nu_T$ are respectively the fluid and turbulence viscosities, $U$ the mean velocity, $y$ the distance from the wall, $u_F$ the friction velocity, and $R$ the half width of the channel. Eq. (1) is derived from the original form in which $\nu dU/dy - \overline{uv} = u_F ^2 (1-y/R)$ \citep{TnL,Hinze,Schlichting,Tritton,LL,Pope,McComb}, where $u$ and $v$ are respectively the streamwise and spanwise fluctuating velocities, and $\overline{(~)}$ the time average. When considering the no-slip condition at the wall meaning that $y=u=v=0$, we obtain the definition of $u_F^2 = \nu dU/dy$. By assuming that turbulence effectively creates the `extra' viscosity $\nu_T$, we may replace $- \overline{uv}$ with $\nu_T dU/dy$ in the original balance to obtain Eq. (1) \citep{NoteEq1}. Now we attempt from Eq. (1) to reach the log-law MVP which has the dimensionless form
\begin{eqnarray}
U_+ = (1/\kappa) ln(y_+) + B,
\end{eqnarray}
where $U_+ = U/u_F$ and $y_+ = yu_F/\nu$ as well as $\kappa$ the von K\'arm\'an constant and $B$ an offset constant \citep{TnL,Hinze,Schlichting,Tritton,LL,Pope,McComb}. Physically, the log law exists in the range of $y$ where it is not only far enough away from the wall in order for $\nu_T \gg \nu$ but also far away from the channel centerline ($y=R$) so that $y \ll R$. Therefore, from Eq. (1), this implies
\begin{eqnarray}
\nu_TdU/dy = u_F ^2.
\end{eqnarray}
It is then clear from Eq. (3) that we can obtain Eq. (2) by taking
\begin{eqnarray}
\nu_T = \kappa yu_F.
\end{eqnarray}
Following Eq. (4), it seems appropriate to define a turbulent Reynolds number for channel flow as $Re_T = \nu_{T,max}/\nu$, where the maximum turbulent diffusivity $\nu_{T,max}$ should be proportional to $\kappa R u_F$. This signifies the usage of frictional Reynolds number $Re_F = Ru_F/\nu$ for characterizing the wall-bounded flow. Based on Eq. (4), we assume the minimum turbulent diffusivity $\nu_{T,min}$ as
\begin{eqnarray}
\ell_T u_T = \kappa \delta_F u_F,
\end{eqnarray}
where $\ell_T$ and $u_T$ are respectively the length and velocity scales of the minimum turbulent fluctuation, and $\delta_F = \nu/u_F$ is the friction length. This implies $\kappa$ as the ratio of $\nu_{T,min} = \ell_Tu_T$ to the frictional diffusivity $\delta_Fu_F$ or simply $\nu$.

Here we attempt to simulate the dimensionless MVP that features the log-law layer in Eq. (2). We consider the progressive linear approximation
\begin{eqnarray}
U_{+,y_+ + \Delta y_+} = U_{+,y_+} + (dU_+/dy_+)_{y_+} \Delta y_+,
\end{eqnarray}
where $U_{+,y_+ + \Delta y_+}$ represents the dimensionless velocity $U_+$ at the dimensionless position $y_+ + \Delta y_+$, similarly $U_{+,y_+}$ is at $y_+$, and the dimensionless velocity gradient or shear stress $(dU_+/dy_+)_{y_+}$ at $y_+$ comes solely from the contribution of multiscale turbulent motions. We apply the fixed-end standing waves with even harmonic modes in the channel (please see Fig. 1a for example), and show that the wave envelope of superposition of these standing waves, or presumably the superposition of multiscale turbulent shear stresses, is capable of constructing the shear stress profile $dU_+/dy_+$ in terms of $y/R$ across the channel (please see \citep{Pope} or the `sum of envelopes' in Fig. 1b here for example). Why nature favors only such even-mode harmonic waves is open for discovery elsewhere. Nevertheless, we can clearly determine whether the select standing waves work for the current purpose or not by carefully examining their wave superposition and the associated wave envelope.

We observe that the fixed-end standing waves with even harmonic modes (Fig. 1a) are symmetric with respect to $y/R=1$, consistent with the known shear stress profile in channel flow (please see \cite{Pope} for example). The equation of motion of these standing waves is $A sin(ky) cos(\omega t)$, where $A$ is any given amplitude regardless of the wavenumbers $k$, the angular frequency $\omega = 2\pi/\tau$ where $\tau$ is the associated turbulent time scale, and $t$ the time. $k = n\pi/D$, where only the even harmonic modes $n = 2, 4, 6, ...$ are used hereafter for the log-law MVP except for comparison, and the channel width $D = 2R$. Here we take $R = \pi$. Turbulent length scales are defined as $\ell = \lambda/4\pi = 1/2k$ where $\lambda$ is the wavelength. Therefore, the maximum turbulent length scale $\ell_{max} = 1/2$ corresponding to the minimum value of $n$, that is, $n_{min} = 2$. The larger the given maximum value of $n$, $n_{max}$, the greater depth the turbulent activity. This is because the resolving length scale $\ell_{min} = \ell_T = 1/n_{max}$.

Here we further simplify the wave motion from $A sin(ky) cos(\omega t)$ to $sin(ky)$ with the following arguments. We infer from Eq. (4) that $\nu_T$ could stop growing when $y_+ \geq Re_T$. For simplicity, we assume that there exists a constant $\nu_{T,max}$ for $y_+ \geq Re_T$. Note that $\nu_T$ at a scale $\ell'$ represents the integral of all viscous effects from all $\ell < \ell'$ \citep{McComb}. Since the energy rate $u^3/\ell$ is conserved across the turbulent scales in the inertial subrange, $\nu_T$ should be at least approximately a constant. Therefore, it might not be far away from plausible to associated the constant $\nu_{T,max}$ with the inertial subrange of turbulence. This implies that the turbulent scales in the log-law regime are smaller than those in the inertial subrange where $u \sim \ell^{1/3}$. In other words, all turbulent scales in such regime are affected by viscosity, suggesting a linear scaling of velocity with size \citep{K41,LL}. That is, $u \sim \ell$, implying $\omega = 2\pi/\tau \sim du/dy \sim c = const.$ regardless of $\ell$ within the log-law regime. The wave motion may then be simplified as $A |cos(ct)| sin(ky)$. Note that the absolute value of $cos(ct)$ is taken to avoid the change of sign in the wave superposition. This is to be consistent with the fact that the turbulent stress $-\overline{uv} > 0$ is always true where $dU/dy > 0$, and vice versa. Moreover, since the maximum of $dU_+/dy_+$ is bounded by unity at the wall, according to Eq. (1) with both $y=0$ and $\nu_T=0$, the wave superposition is always normalized by its own maximum peak value before being applied to Eq. (6). Consequently, $A |cos(ct)|>0$ which has no influence on the normalized result can thus be simply neglected. The trivial result from $A |cos(ct)|=0$ is not considered here either. Therefore, the wave motion of interest reduces to the simple harmonic sine wave $sin(ky)$.

We now can approximate $dU_+/dy_+$ with the following:
\begin{widetext}
\begin{equation}
d{U_ + }/d{y_ + } \approx \mathop {{\rm{envelope}}}\limits_{maxima} \left\{ {{{\sum\limits_{n = 2,4,6,...}^{{n_{max}}} s in\left( {\frac{{n\pi }}{D}y} \right)} \mathord{\left/
 {\vphantom {{\sum\limits_{n = 2,4,6,...}^{{n_{max}}} s in\left( {\frac{{n\pi }}{D}y} \right)} {\mathop {\max }\limits_y \left\{ {\sum\limits_{n = 2,4,6,...}^{{n_{max}}} s in\left( {\frac{{n\pi }}{D}y} \right)} \right\}}}} \right.
 \kern-\nulldelimiterspace} {\mathop {\max }\limits_y \left\{ {\sum\limits_{n = 2,4,6,...}^{{n_{max}}} s in\left( {\frac{{n\pi }}{D}y} \right)} \right\}}}} \right\},
\end{equation}
\end{widetext}
where $envelope\{\}$ is to determine the wave envelope of the wave superposition using spline interpolation over the local maxima of the superposition and $max\{\}$ is to find the maximum peak value of the superposition with respect to $y$. Note here $y$ is from 0 to $D$. Fig. 1(b) shows the wave superposition and its associated wave envelopes obtained using Eq. (7) with $n_{max} = 50$ across the channel. Since the free-end standing waves with odd harmonic modes are also symmetric about $y/R = 0$, their superposition whose simplified motion has the form of $cos(ky)$ is also shown in Fig. 1(b) for comparison. It is clear that the latter cannot generate the shear stress profile of interest. It should be noted as follows. From Eq. (1) with $\nu_T = 0$, we have $dU_+/dy_+ = 1 - y/R$. There are two associated boundary conditions to be satisfied here, $dU_+/dy_+ = 1$ at $y/R = 0$ and $dU_+/dy_+ = 0$ at $y/R = 1$. The upper and lower envelopes shown in Fig. 1(b) are determined respectively over local maxima and minima. We observe that the upper envelope reaches a positive finite nonzero value when approaching $y/R = 1$ (Fig. 1c), while the lower envelope reaches a negative value of almost similar magnitude. We therefore take the sum of upper and lower envelopes as the wave envelope to be used. Moreover, that the wave envelope overshoots the highest peak of wave superposition when approaching $y/R = 0$, is also settled by taking the amplitude of the wave envelope to be unity for $y/R$ between 0 and such peak.

Fig. 1(e) shows the wave envelopes or the equivalent $dU_+/dy_+$ from Eq. (7) as a function of $y/R$ \citep{NoteRes}. To determine MVP with Eq. (6), we need to know how to convert $\Delta y$ to $\Delta y_+$. By taking $u_T = u_F$ in Eq. (5), we have $\ell_T = \kappa \delta_F$. Owing to $\Delta y_+ = \Delta y/\delta_F$, we consequently have 
\begin{eqnarray}
\Delta y_+ = \kappa\Delta y/\ell_T.
\end{eqnarray}
Fig. 1(f) shows the semi-log plot of the MVP's from Eq. (6-8) with the data in Fig. 1(e), where $\kappa = 0.42$ is used. It is clear that these MVP's reveal a self-similar log-law nature. Also shown for comparison in Fig. 1(f) are the MVP's with the same $n_{max}$, but including all modes (from $n = 1$ to $n_{max}$) of the current harmonic waves. Although the log-law nature is retained (Fig. 1f), it is obvious that $dU_+/dy_+$ does not reduce to zero when approaching $y/R = 1$ in those MVP's containing odd harmonic modes.

\begin{figure}
\includegraphics[width=0.4\textwidth, clip, trim=233 138 233 72, angle=0]{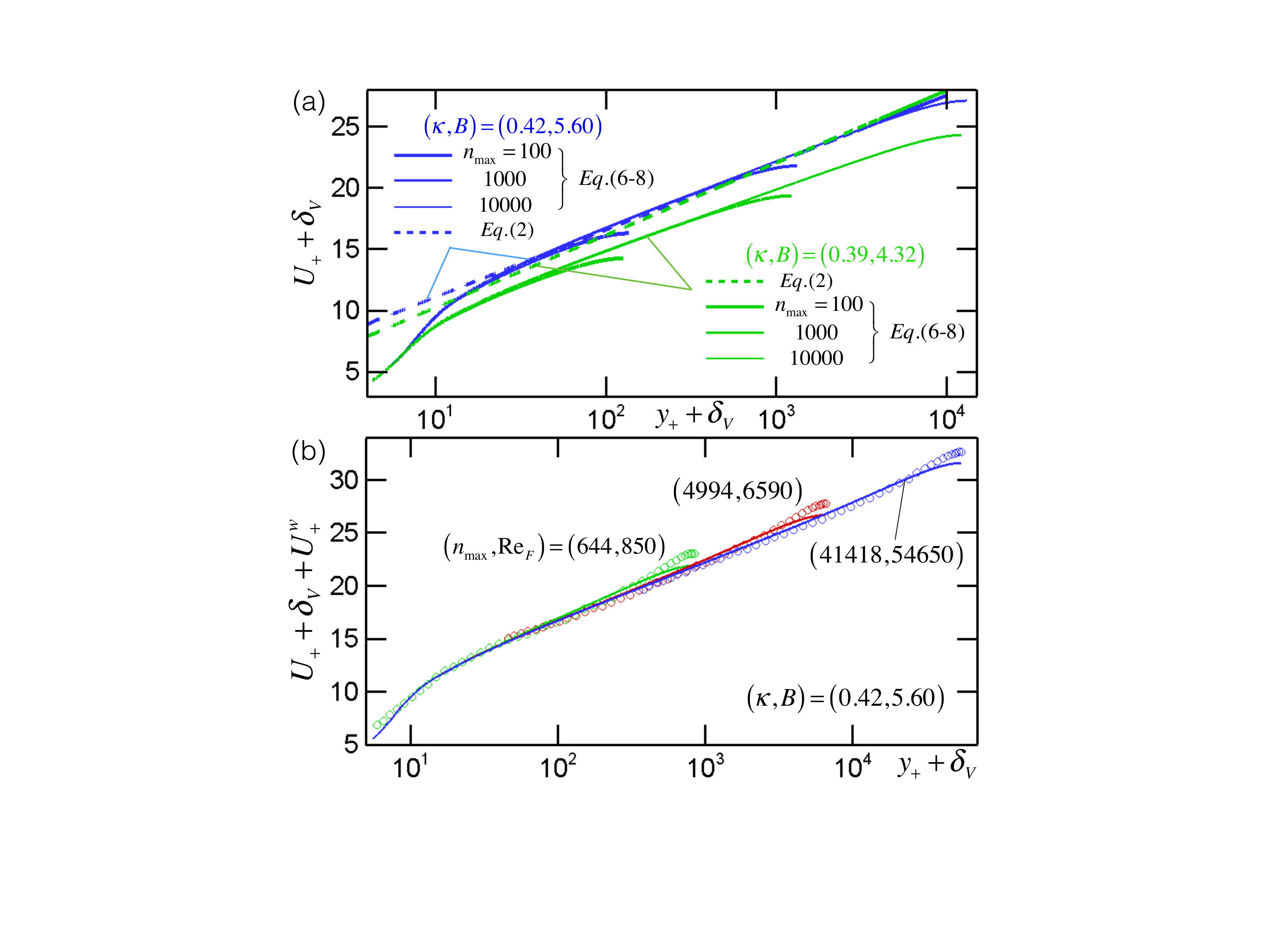}
\caption{\label{fig:epsart} 
(a) MVP's (solid lines) from Eq. (6-8) using wave superposition with the recognized experimental values of $\kappa$ and $B$ from pipe flows \citep{ZS} (blue) and boundary layer flow \citep{MelbourneLogLaw} (green), together with the associated log laws (dash lines) from Eq. (2). Note that the effect of $\delta_V$ is considered by adding $\delta_V$ to both axes. (b) Comparison of the simulated MVP's plus the wake $U_+ ^w$ in Eq. 9 (curves) with the experimental counterparts (circles) from the Princeton Superpipe \citep{ZS} for the same conditions of $Re_F$ in terms of $n_{max}$. Note $Re_F = \kappa \pi n_{max}$.
}
\end{figure}

To make a direct comparison between the simulated MVP and the log law in Eq. (2), we need to consider the existence of the viscous sublayer of thickness $\delta_V$ right next to the wall boundary. Within $\delta_V$, turbulence is assumed to play no significant roles. From Eq. (1), within $\delta_V$ where $\nu \gg \nu_T$ and $y \ll R$, we have a linear profile $U_+ = y_+$. For simplicity, we thus assume the mutual independence between such viscous layer and the simulated MVP together with $\delta_V = B$ \citep{NoteB}. That is to say, $y_+=0$ considered here in Eq. (6) represents a virtual wall for turbulent motion, without awareness of the presence of $\delta_V$. Therefore, to account for the effect of $\delta_V$, we need only to add $\delta_V = B$ to both the position and velocity axes.

Fig. 2(a) shows MVP's derived from Eq. (6-8) with the known experiment results of ($\kappa,B$) in turbulent wall-bounded flows, respectively ($0.39,4.32$) from the Melbourne wind tunnel \citep{MelbourneLogLaw} and ($0.42,5.60$) from the Princeton Superpipe \citep{ZS}. The log-law slopes seem to be consistent between the simulated MVP's and Eq. (2) with both pairs of ($\kappa,B$). The standing-wave MVP's show overall consistency with the result of Superpipe. However, there is an offset between the boundary layer data \citep{MelbourneLogLaw} and MVP's here. Such offset is much large as compared with the difference between two experimental log laws. This indicates that the result from Eq. (6-8) is rather sensitive to the change of ($\kappa,B$).

The region deviates from the log-law layer at large $y_+$ is called wake $U_+ ^w$ \citep{TnL,Pope}. We approximate $U_+ ^w$ by integrating Eq. (1) using $\nu_{T,max} \approx \kappa Ru_F \gg \nu$ with $U_+ ^w = 0$ at $y/R = 0$. That is,
\begin{eqnarray} 
U_+ ^w = (2y_+/R_+ - (y_+/R_+)^2)/2\kappa.
\end{eqnarray}
$U_+ ^w$ can be added to the simulated MVP, since they shares the $y_+$ coordinate. Fig. 2(b) shows the comparison between the Superpipe data \citep{ZS} and MVP's here plus $U_+ ^w$. However, the appreciable differences in the wake regions (Fig. 2b) show the poor approximation of strictly constant $\nu_{T,max}$.

\begin{figure}
\includegraphics[width=0.43\textwidth, clip, trim=115 100 165 45, angle=0]{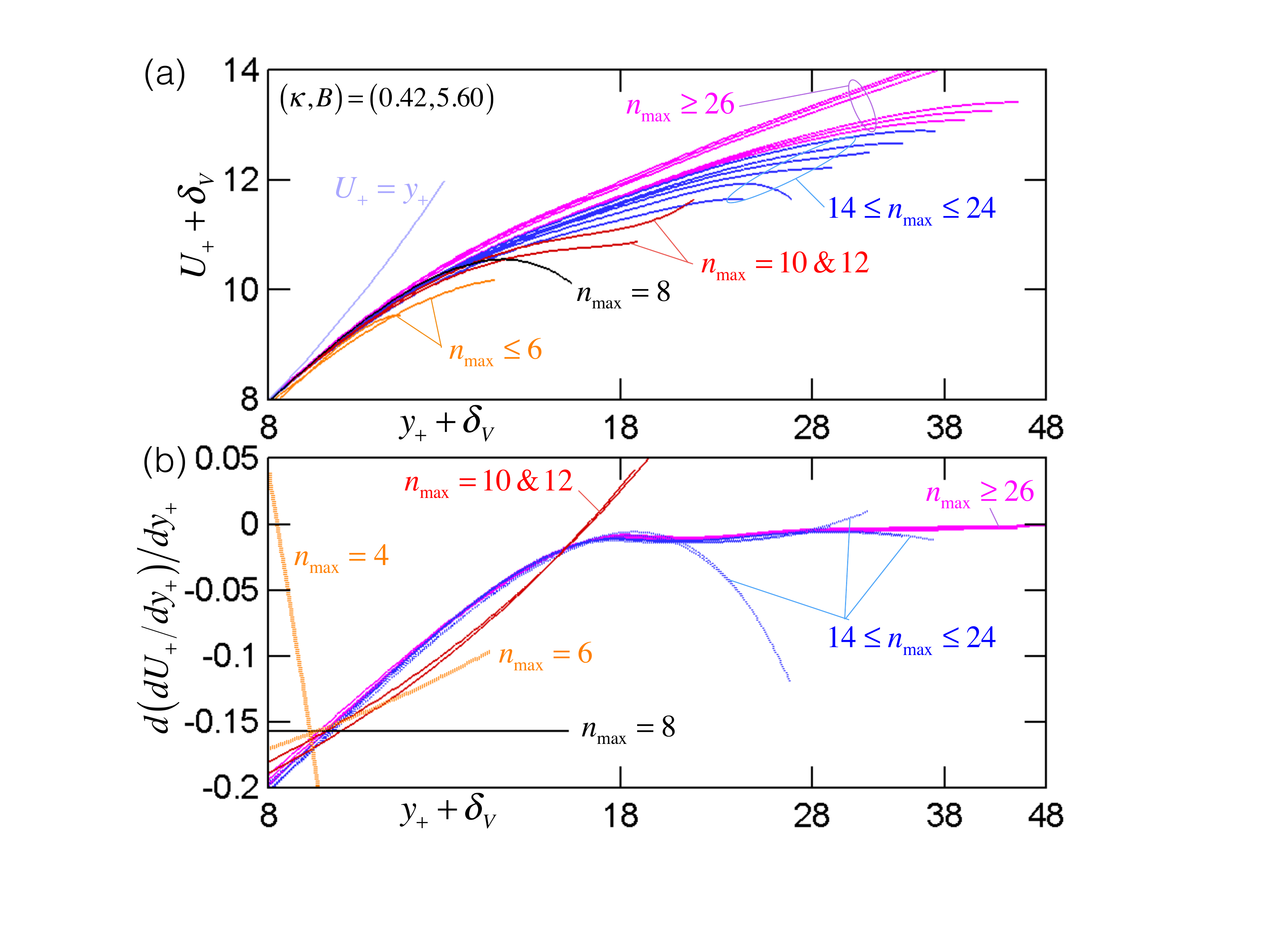}
\caption{\label{fig:epsart} 
(a) The effect of $n_{max}$ on MVP's with (b) their associated curvature profiles, revealing the laminar-like profile with constant negative curvature at $n_{max}=8$, the rise of instability with a point of inflection at $n_{max} = 10,12$, the self-similar curvature profiles when $n_{max} \geq 26$. The linear profile $U_+ = y_+$ is also shown in (a) for comparison.
}
\end{figure}

Interests may also be drawn to the discussion of flow transition. To this end, we make a bold assumption that Eq. (1) is valid at very small Reynolds numbers. In other words, we imagine that even those turbulent patches or puffs (please see \citep{Mullin} for example) which are locally existent could be quasi stationary, when they are convected by steady mean flow. We then postulate, according to the finding here, the turbulent shear stresses in terms of $sin(ky)$ remain capable of dominating MVP at low Reynolds numbers. Now we are interested in seeing whether there exists a distinguishable change in the simulated MVP when $n_{max}$ is gradually reduced. To recognize the sign of a possible transition \citep{Hinze}, we may see if there exists a point of inflection with null curvature in the simulated MVP's at lower $n_{max}$. Fig. 3(a,b) show respectively the simulated MVP's and the associated curvature profiles $d(dU_+/dy_+)/dy_+$ for $n_{max}$ reducing from 1000 to 2. We see that MVP's with roughly $n_{max} \geq 26$ reveal a self-similar curvature distribution (Fig. 3b). However, the curvature of MVP starts to vary when $n_{max} < 26$. Then it jumps from a rather random state where $14 \leq n_{max} \leq 24$ to a laminar-like state at $n_{max} = 8$ with a constant negative curvature. Such transition is signified by the appearance of a point of inflection at $n_{max} = 10,12$. For $n_{max} \leq 6$, observation becomes difficult. This might imply that Eq. (7) is valid only for $n_{max} \geq 8$. We hope that these preliminary results could motivate new ideas in the studies of flow transition.

The even harmonic modes of fixed-end standing waves render their amplitudes and wavelengths respectively into the dimensionless shear stresses and length scales of stationary wall-bounded turbulent flow. The wave envelope of their superposition thus reproduces the log-law MVP, providing new diagnostics for investigating the wall-turbulence interaction. The result also suggests the need for reconsideration of the existent theories. Moreover, when gradually reducing the harmonic modes, we also see the possibility for such application to dealing with the flow transition. We anticipate that diagnostics with harmonic standing waves might lead to enhance understanding of confined turbulence.

The author is grateful to the Max Planck Society and the Okinawa Institute of Science and Technology for their support.


\begin{thebibliography}{24}%
\makeatletter
\providecommand \@ifxundefined [1]{%
 \@ifx{#1\undefined}
}%
\providecommand \@ifnum [1]{%
 \ifnum #1\expandafter \@firstoftwo
 \else \expandafter \@secondoftwo
 \fi
}%
\providecommand \@ifx [1]{%
 \ifx #1\expandafter \@firstoftwo
 \else \expandafter \@secondoftwo
 \fi
}%
\providecommand \natexlab [1]{#1}%
\providecommand \enquote  [1]{``#1''}%
\providecommand \bibnamefont  [1]{#1}%
\providecommand \bibfnamefont [1]{#1}%
\providecommand \citenamefont [1]{#1}%
\providecommand \href@noop [0]{\@secondoftwo}%
\providecommand \href [0]{\begingroup \@sanitize@url \@href}%
\providecommand \@href[1]{\@@startlink{#1}\@@href}%
\providecommand \@@href[1]{\endgroup#1\@@endlink}%
\providecommand \@sanitize@url [0]{\catcode `\\12\catcode `\$12\catcode
  `\&12\catcode `\#12\catcode `\^12\catcode `\_12\catcode `\%12\relax}%
\providecommand \@@startlink[1]{}%
\providecommand \@@endlink[0]{}%
\providecommand \url  [0]{\begingroup\@sanitize@url \@url }%
\providecommand \@url [1]{\endgroup\@href {#1}{\urlprefix }}%
\providecommand \urlprefix  [0]{URL }%
\providecommand \Eprint [0]{\href }%
\providecommand \doibase [0]{http://dx.doi.org/}%
\providecommand \selectlanguage [0]{\@gobble}%
\providecommand \bibinfo  [0]{\@secondoftwo}%
\providecommand \bibfield  [0]{\@secondoftwo}%
\providecommand \translation [1]{[#1]}%
\providecommand \BibitemOpen [0]{}%
\providecommand \bibitemStop [0]{}%
\providecommand \bibitemNoStop [0]{.\EOS\space}%
\providecommand \EOS [0]{\spacefactor3000\relax}%
\providecommand \BibitemShut  [1]{\csname bibitem#1\endcsname}%
\let\auto@bib@innerbib\@empty
\bibitem [{\citenamefont {Tennekes}\ and\ \citenamefont {Lumley}(1972)}]{TnL}%
  \BibitemOpen
  \bibfield  {author} {\bibinfo {author} {\bibfnamefont {H.}~\bibnamefont
  {Tennekes}}\ and\ \bibinfo {author} {\bibfnamefont {J.~L.}\ \bibnamefont
  {Lumley}},\ }\href@noop {} {\emph {\bibinfo {title} {A First Course in
  Turbulence}}}\ (\bibinfo  {publisher} {MIT Press},\ \bibinfo {year}
  {1972})\BibitemShut {NoStop}%
\bibitem [{\citenamefont {Hinze}(1975)}]{Hinze}%
  \BibitemOpen
  \bibfield  {author} {\bibinfo {author} {\bibfnamefont {J.~O.}\ \bibnamefont
  {Hinze}},\ }\href@noop {} {\emph {\bibinfo {title} {Turbulence}}},\ \bibinfo
  {edition} {2nd}\ ed.\ (\bibinfo  {publisher} {McGraw-Hill},\ \bibinfo {year}
  {1975})\BibitemShut {NoStop}%
\bibitem [{\citenamefont {Schlichting}(1979)}]{Schlichting}%
  \BibitemOpen
  \bibfield  {author} {\bibinfo {author} {\bibfnamefont {H.}~\bibnamefont
  {Schlichting}},\ }\href@noop {} {\emph {\bibinfo {title} {Boundary-Layer
  Theory}}},\ \bibinfo {edition} {7th}\ ed.\ (\bibinfo  {publisher}
  {McGraw-Hill},\ \bibinfo {year} {1979})\BibitemShut {NoStop}%
\bibitem [{\citenamefont {Tritton}(1988)}]{Tritton}%
  \BibitemOpen
  \bibfield  {author} {\bibinfo {author} {\bibfnamefont {D.~J.}\ \bibnamefont
  {Tritton}},\ }\href@noop {} {\emph {\bibinfo {title} {Physical Fluid
  Dynamics}}},\ \bibinfo {edition} {2nd}\ ed.\ (\bibinfo  {publisher} {Oxford
  University Press},\ \bibinfo {year} {1988})\BibitemShut {NoStop}%
\bibitem [{\citenamefont {Landau}\ and\ \citenamefont {Lifshitz}(1987)}]{LL}%
  \BibitemOpen
  \bibfield  {author} {\bibinfo {author} {\bibfnamefont {L.}~\bibnamefont
  {Landau}}\ and\ \bibinfo {author} {\bibfnamefont {E.}~\bibnamefont
  {Lifshitz}},\ }\href@noop {} {\emph {\bibinfo {title} {Fluid Mechanics}}},\
  \bibinfo {edition} {2nd}\ ed.\ (\bibinfo  {publisher} {Elsevier},\ \bibinfo
  {year} {1987})\BibitemShut {NoStop}%
\bibitem [{\citenamefont {Pope}(2000)}]{Pope}%
  \BibitemOpen
  \bibfield  {author} {\bibinfo {author} {\bibfnamefont {S.~B.}\ \bibnamefont
  {Pope}},\ }\href@noop {} {\emph {\bibinfo {title} {Turbulent Flows}}}\
  (\bibinfo  {publisher} {Cambridge University Press},\ \bibinfo {year}
  {2000})\BibitemShut {NoStop}%
\bibitem [{\citenamefont {McComb}(1990)}]{McComb}%
  \BibitemOpen
  \bibfield  {author} {\bibinfo {author} {\bibfnamefont {W.~D.}\ \bibnamefont
  {McComb}},\ }\href@noop {} {\emph {\bibinfo {title} {The Physics of Fluid
  Turbulence}}}\ (\bibinfo  {publisher} {Oxford University Press},\ \bibinfo
  {year} {1990})\BibitemShut {NoStop}%
\bibitem [{\citenamefont {McWilliams}(2006)}]{GFD}%
  \BibitemOpen
  \bibfield  {author} {\bibinfo {author} {\bibfnamefont {J.~C.}\ \bibnamefont
  {McWilliams}},\ }\href@noop {} {\emph {\bibinfo {title} {Fundamentals of
  Geophysical Fluid Mechanics}}}\ (\bibinfo  {publisher} {Cambridge University
  Press},\ \bibinfo {year} {2006})\BibitemShut {NoStop}%
\bibitem [{\citenamefont {Wyngaard}(2010)}]{AFD}%
  \BibitemOpen
  \bibfield  {author} {\bibinfo {author} {\bibfnamefont {J.~C.}\ \bibnamefont
  {Wyngaard}},\ }\href@noop {} {\emph {\bibinfo {title} {Turbulence in the
  atmosphere}}}\ (\bibinfo  {publisher} {Cambridge University Press},\ \bibinfo
  {year} {2010})\BibitemShut {NoStop}%
\bibitem [{\citenamefont {Gioia}\ \emph {et~al.}(2010)\citenamefont {Gioia},
  \citenamefont {Guttenberg}, \citenamefont {Goldenfeld},\ and\ \citenamefont
  {Chakraborty}}]{GG2010}%
  \BibitemOpen
  \bibfield  {author} {\bibinfo {author} {\bibfnamefont {G.}~\bibnamefont
  {Gioia}}, \bibinfo {author} {\bibfnamefont {N.}~\bibnamefont {Guttenberg}},
  \bibinfo {author} {\bibfnamefont {N.}~\bibnamefont {Goldenfeld}}, \ and\
  \bibinfo {author} {\bibfnamefont {P.}~\bibnamefont {Chakraborty}},\
  }\href@noop {} {\bibfield  {journal} {\bibinfo  {journal} {Phys. Rev. Lett.}\
  }\textbf {\bibinfo {volume} {105}} (\bibinfo {year} {2010})}\BibitemShut
  {NoStop}%
\bibitem [{\citenamefont {Marusic}\ \emph
  {et~al.}(2010{\natexlab{a}})\citenamefont {Marusic}, \citenamefont {McKeon},
  \citenamefont {Monkewitz}, \citenamefont {Nagib}, \citenamefont {Smits},\
  and\ \citenamefont {Sreenivasan}}]{WBF2010}%
  \BibitemOpen
  \bibfield  {author} {\bibinfo {author} {\bibfnamefont {I.}~\bibnamefont
  {Marusic}}, \bibinfo {author} {\bibfnamefont {B.~J.}\ \bibnamefont {McKeon}},
  \bibinfo {author} {\bibfnamefont {P.~A.}\ \bibnamefont {Monkewitz}}, \bibinfo
  {author} {\bibfnamefont {H.~M.}\ \bibnamefont {Nagib}}, \bibinfo {author}
  {\bibfnamefont {A.~J.}\ \bibnamefont {Smits}}, \ and\ \bibinfo {author}
  {\bibfnamefont {K.~R.}\ \bibnamefont {Sreenivasan}},\ }\href@noop {}
  {\bibfield  {journal} {\bibinfo  {journal} {Phys. Fluids}\ }\textbf {\bibinfo
  {volume} {22}},\ \bibinfo {pages} {065103} (\bibinfo {year}
  {2010}{\natexlab{a}})}\BibitemShut {NoStop}%
\bibitem [{\citenamefont {Marusic}\ \emph
  {et~al.}(2010{\natexlab{b}})\citenamefont {Marusic}, \citenamefont {Mathis},\
  and\ \citenamefont {Hutchins}}]{WBFScience}%
  \BibitemOpen
  \bibfield  {author} {\bibinfo {author} {\bibfnamefont {I.}~\bibnamefont
  {Marusic}}, \bibinfo {author} {\bibfnamefont {R.}~\bibnamefont {Mathis}}, \
  and\ \bibinfo {author} {\bibfnamefont {N.}~\bibnamefont {Hutchins}},\
  }\href@noop {} {\bibfield  {journal} {\bibinfo  {journal} {Science}\ }\textbf
  {\bibinfo {volume} {329}},\ \bibinfo {pages} {193} (\bibinfo {year}
  {2010}{\natexlab{b}})}\BibitemShut {NoStop}%
\bibitem [{\citenamefont {Barkley}\ \emph {et~al.}(2015)\citenamefont
  {Barkley}, \citenamefont {Song}, \citenamefont {Mukund}, \citenamefont
  {Lemoult}, \citenamefont {Avila},\ and\ \citenamefont {Hof}}]{WBFNature}%
  \BibitemOpen
  \bibfield  {author} {\bibinfo {author} {\bibfnamefont {D.}~\bibnamefont
  {Barkley}}, \bibinfo {author} {\bibfnamefont {B.}~\bibnamefont {Song}},
  \bibinfo {author} {\bibfnamefont {V.}~\bibnamefont {Mukund}}, \bibinfo
  {author} {\bibfnamefont {G.}~\bibnamefont {Lemoult}}, \bibinfo {author}
  {\bibfnamefont {M.}~\bibnamefont {Avila}}, \ and\ \bibinfo {author}
  {\bibfnamefont {B.}~\bibnamefont {Hof}},\ }\href@noop {} {\bibfield
  {journal} {\bibinfo  {journal} {Nature}\ }\textbf {\bibinfo {volume} {526}},\
  \bibinfo {pages} {550} (\bibinfo {year} {2015})}\BibitemShut {NoStop}%
\bibitem [{\citenamefont {Faisst}\ and\ \citenamefont
  {Eckhardt}(2003)}]{TWPRL}%
  \BibitemOpen
  \bibfield  {author} {\bibinfo {author} {\bibfnamefont {H.}~\bibnamefont
  {Faisst}}\ and\ \bibinfo {author} {\bibfnamefont {B.}~\bibnamefont
  {Eckhardt}},\ }\href@noop {} {\bibfield  {journal} {\bibinfo  {journal}
  {Phys. Rev. Lett.}\ }\textbf {\bibinfo {volume} {91}},\ \bibinfo {pages}
  {224502} (\bibinfo {year} {2003})}\BibitemShut {NoStop}%
\bibitem [{\citenamefont {Hof}\ \emph {et~al.}(2004)\citenamefont {Hof},
  \citenamefont {van Doorne}, \citenamefont {Westerweel}, \citenamefont
  {Nieuwstadt}, \citenamefont {Faisst}, \citenamefont {Eckhardt}, \citenamefont
  {Wedin}, \citenamefont {Kerswell},\ and\ \citenamefont
  {Waleffe}}]{TWScience}%
  \BibitemOpen
  \bibfield  {author} {\bibinfo {author} {\bibfnamefont {B.}~\bibnamefont
  {Hof}}, \bibinfo {author} {\bibfnamefont {C.~W.~H.}\ \bibnamefont {van
  Doorne}}, \bibinfo {author} {\bibfnamefont {J.}~\bibnamefont {Westerweel}},
  \bibinfo {author} {\bibfnamefont {F.~T.~M.}\ \bibnamefont {Nieuwstadt}},
  \bibinfo {author} {\bibfnamefont {H.}~\bibnamefont {Faisst}}, \bibinfo
  {author} {\bibfnamefont {B.}~\bibnamefont {Eckhardt}}, \bibinfo {author}
  {\bibfnamefont {H.}~\bibnamefont {Wedin}}, \bibinfo {author} {\bibfnamefont
  {R.~R.}\ \bibnamefont {Kerswell}}, \ and\ \bibinfo {author} {\bibfnamefont
  {F.}~\bibnamefont {Waleffe}},\ }\href@noop {} {\bibfield  {journal} {\bibinfo
   {journal} {Science}\ }\textbf {\bibinfo {volume} {305}},\ \bibinfo {pages}
  {1594} (\bibinfo {year} {2004})}\BibitemShut {NoStop}%
\bibitem [{\citenamefont {Duggleby}\ \emph {et~al.}(2009)\citenamefont
  {Duggleby}, \citenamefont {Ball},\ and\ \citenamefont {Schwaenen}}]{TW}%
  \BibitemOpen
  \bibfield  {author} {\bibinfo {author} {\bibfnamefont {A.}~\bibnamefont
  {Duggleby}}, \bibinfo {author} {\bibfnamefont {K.~S.}\ \bibnamefont {Ball}},
  \ and\ \bibinfo {author} {\bibfnamefont {M.}~\bibnamefont {Schwaenen}},\
  }\href@noop {} {\bibfield  {journal} {\bibinfo  {journal} {Phil. Trans. R.
  Soc. A}\ }\textbf {\bibinfo {volume} {367}},\ \bibinfo {pages} {473}
  (\bibinfo {year} {2009})}\BibitemShut {NoStop}%
\bibitem [{\citenamefont {Boussinesq}(1877)}]{Boussinesq}%
  \BibitemOpen
  \bibfield  {author} {\bibinfo {author} {\bibfnamefont {J.}~\bibnamefont
  {Boussinesq}},\ }\href@noop {} {\bibfield  {journal} {\bibinfo  {journal}
  {M{\'e}moires pr{\'e}sent{\'e}s par divers savants {\`a} l'Acad{\'e}mie des
  Sciences}\ }\textbf {\bibinfo {volume} {23}},\ \bibinfo {pages} {1} (\bibinfo
  {year} {1877})}\BibitemShut {NoStop}%
\bibitem [{Not()}]{NoteEq1}%
  \BibitemOpen
  \href@noop {} {}\bibinfo {note} {On the left hand side of Eq. (1), the first
  term comes from the contribution of the viscous effect and the second term
  represents the contribution from the multiscale turbulent motions. The sum of
  these two terms, which gives the shear stress profile or MVP for wall-bounded
  flow, equals to the linear total stress relation on the right hand side of
  Eq. (1).}\BibitemShut {Stop}%
\bibitem [{\citenamefont {Kolmogorov}(1991)}]{K41}%
  \BibitemOpen
  \bibfield  {author} {\bibinfo {author} {\bibfnamefont {A.~N.}\ \bibnamefont
  {Kolmogorov}},\ }\href@noop {} {\bibfield  {journal} {\bibinfo  {journal}
  {Proc. R. Soc. Lond. A}\ }\textbf {\bibinfo {volume} {434}},\ \bibinfo
  {pages} {9} (\bibinfo {year} {1991})}\BibitemShut {NoStop}%
\bibitem [{()}]{NoteRes}%
  \BibitemOpen
  \href@noop {} {}\bibinfo {note} {Since $\ell_T/R = 1/\pi n_{max}$, the
  resolution $\Delta y/R$ in determination of the wave superposition requires
  that $\Delta y/R \leq 1/\pi n_{max}$. However, we found that the results
  start to be becoming independent of $\Delta y/R$ only when $\Delta y/R \leq
  1/4\pi n_{max}$, which is the resolution used here.}\BibitemShut {Stop}%
\bibitem [{\citenamefont {Zagarola}\ and\ \citenamefont {Smits}(1998)}]{ZS}%
  \BibitemOpen
  \bibfield  {author} {\bibinfo {author} {\bibfnamefont {M.~V.}\ \bibnamefont
  {Zagarola}}\ and\ \bibinfo {author} {\bibfnamefont {A.~J.}\ \bibnamefont
  {Smits}},\ }\href@noop {} {\bibfield  {journal} {\bibinfo  {journal} {J.
  Fluid Mech.}\ }\textbf {\bibinfo {volume} {373}},\ \bibinfo {pages} {33}
  (\bibinfo {year} {1998})}\BibitemShut {NoStop}%
\bibitem [{\citenamefont {Marusic}\ \emph {et~al.}(2013)\citenamefont
  {Marusic}, \citenamefont {Monty}, \citenamefont {Hultmark},\ and\
  \citenamefont {Smits}}]{MelbourneLogLaw}%
  \BibitemOpen
  \bibfield  {author} {\bibinfo {author} {\bibfnamefont {I.}~\bibnamefont
  {Marusic}}, \bibinfo {author} {\bibfnamefont {J.~P.}\ \bibnamefont {Monty}},
  \bibinfo {author} {\bibfnamefont {M.}~\bibnamefont {Hultmark}}, \ and\
  \bibinfo {author} {\bibfnamefont {A.~J.}\ \bibnamefont {Smits}},\ }\href@noop
  {} {\bibfield  {journal} {\bibinfo  {journal} {J. Fluid Mech.}\ }\textbf
  {\bibinfo {volume} {716}},\ \bibinfo {pages} {R3} (\bibinfo {year}
  {2013})}\BibitemShut {NoStop}%
\bibitem [{\citenamefont {.}()}]{NoteB}%
  \BibitemOpen
  \href@noop {} {}\bibinfo {note} {Physically, $B$ in Eq. (2) or similarly as the constant of
  integration from Eq. (3) accounts for the existence of the viscous layer of
  thickness $\delta_V$ between the wall and the log-law MVP
  \citep{LL}.}\BibitemShut {Stop}%
\bibitem [{\citenamefont {Willis}\ \emph {et~al.}(2008)\citenamefont {Willis},
  \citenamefont {Peixinho}, \citenamefont {Kerswell},\ and\ \citenamefont
  {Mullin}}]{Mullin}%
  \BibitemOpen
  \bibfield  {author} {\bibinfo {author} {\bibfnamefont {A.~P.}\ \bibnamefont
  {Willis}}, \bibinfo {author} {\bibfnamefont {J.}~\bibnamefont {Peixinho}},
  \bibinfo {author} {\bibfnamefont {R.~R.}\ \bibnamefont {Kerswell}}, \ and\
  \bibinfo {author} {\bibfnamefont {T.}~\bibnamefont {Mullin}},\ }\href@noop {}
  {\bibfield  {journal} {\bibinfo  {journal} {Phil. Trans. R. Soc. A}\ }\textbf
  {\bibinfo {volume} {366}},\ \bibinfo {pages} {2671} (\bibinfo {year}
  {2008})}\BibitemShut {NoStop}%
\end{thebibliography}
%

\end{document}